\documentclass[%
  twocolumn,
  aps,
  prb,
  amsmath,
  amssymb,
  superscriptaddress,
  nofootinbib,
  floatfix
]{revtex4-1}

\usepackage{graphicx}
\usepackage{hyperref}
\usepackage{dcolumn}
\usepackage{bm}
\usepackage{color}

\newcommand{\s}{\sum\limits}

\newcommand{\pa}{\partial}

\newcommand{\be}{\begin{equation}}
\newcommand{\e}{\end{equation}}
\newcommand{\beml}{\begin{subequations}}
\newcommand{\eml}{\end{subequations}}
\newcommand{\beq}{\begin{eqnarray}}
\newcommand{\eq}{\end{eqnarray}}
\newcommand{\ba}{\begin{array}}
\newcommand{\ea}{\end{array}}
\newcommand{\bpm}{\begin{pmatrix}}
\newcommand{\epm}{\end{pmatrix}}
\newcommand{\bc}{\begin{cases}}
\newcommand{\ec}{\end{cases}}
\newcommand{\lt}{\left}
\newcommand{\rt}{\right}
\newcommand{\n}{\nonumber}
\newcommand{\la}{\langle}
\newcommand{\ra}{\rangle}
\newcommand{\ep}{\varepsilon}

\newcommand{\bb}{\boldsymbol}
\newcommand{\h}{^\dagger}
\newcommand{\0}{^{\phantom{\dagger}}}

\DeclareMathOperator{\tr}{Tr}

\DeclareMathOperator{\im}{Im}

\begin{document}

\title{Giant anisotropy of Gilbert damping in a Rashba honeycomb antiferromagnet}

\author{M. Baglai}
\affiliation{Department of Physics and Astronomy, Uppsala University, Box 516, SE-751 20, Uppsala, Sweden}

\author{R. J. Sokolewicz}
\affiliation{Institute for Molecules and Materials, Radboud University Nijmegen, NL-6525 AJ Nijmegen, the Netherlands}

\author{A. Pervishko}
\affiliation{Department of Physics and Astronomy, Uppsala University, Box 516, SE-751 20, Uppsala, Sweden}
\affiliation{ITMO University, Saint Petersburg 197101, Russia}

\author{M. I. Katsnelson}
\affiliation{Institute for Molecules and Materials, Radboud University Nijmegen, NL-6525 AJ Nijmegen, the Netherlands}

\author{O. Eriksson}
\affiliation{Department of Physics and Astronomy, Uppsala University, Box 516, SE-751 20, Uppsala, Sweden}
\affiliation{School of Science and Technology, \"Orebro University, SE-701 82 \"Orebro, Sweden}

\author{D. Yudin}
\affiliation{Skolkovo Institute of Science and Technology, Moscow 121205, Russia}
\affiliation{Department of Physics and Astronomy, Uppsala University, Box 516, SE-751 20, Uppsala, Sweden}

\author{M. Titov}
\affiliation{Institute for Molecules and Materials, Radboud University Nijmegen, NL-6525 AJ Nijmegen, the Netherlands}
\affiliation{ITMO University, Saint Petersburg 197101, Russia}

\date{\today}

\begin{abstract}
Giant Gilbert damping anisotropy is identified as a signature of strong Rashba spin-orbit coupling in a two-dimensional antiferromagnet on a honeycomb lattice. The phenomenon originates in spin-orbit induced splitting of conduction electron subbands that strongly suppresses certain spin-flip processes. As a result, the spin-orbit interaction is shown to support an undamped non-equilibrium dynamical mode that corresponds to an ultrafast in-plane N\'eel vector precession and a constant perpendicular-to-the-plane magnetization. The phenomenon is illustrated on the basis of a two dimensional $s$-$d$ like model. Spin-orbit torques and conductivity are also computed microscopically for this model. Unlike Gilbert damping these quantities are shown to reveal only a weak anisotropy that is limited to the semiconductor regime corresponding to the Fermi energy staying in a close vicinity of antiferromagnetic gap.
\end{abstract}

\maketitle

\section{Introduction}

A gapless character of the spin-wave spectrum in isotropic Heisenberg magnets in two dimensions results in the homogeneity of magnetic ordering being destroyed by thermal fluctuations at any finite temperatures. In contrast, in van der Waals magnets, characterized by intrinsic magnetocrystalline anisotropy that stems from spin-orbit coupling \cite{Lado2017}, an ordered magnetic state can be retained down to a monolayer limit. Two-dimensional (2D) van der Waals magnets are currently experiencing a revived attention \cite{Gong2017,Herrero2017,Burch2018,Tokmachev2018,Gong2019,Novoselov2019,Cortie2019} driven by the prospects of gateable magnetism \cite{Huang2018,Shengwei2018,Wang2018,Deng2018}, a continuing search for Kitaev materials \cite{Nagler2019,Gordon2019} and Majorana fermions \cite{Livanas2019}, topologically driven phenomena \cite{Mokrousov2019} as well as various applications \cite{Herrero2017,Burch2018,Novoselov2019}. The trade-off between quantum confinement, nontrivial topology and long-range magnetic correlations determines their unique magnetoelectronic properties, in particular a tunable tunneling conductance \cite{Wang2018a} and magnetoresistance \cite{Song2018,Klein2018,Kim2018} depending on the number of layers in the sample, as well as long-distance magnon transport \cite{Xing2019}. 

Ferromagnetic thin films have already entered commercial use in hard drives, magnetic field and rotation angle sensors and in similar devices  \cite{Parkin2003,Jogschies2015,Novoselov2019}, while keeping high promises for technologically competitive ultrafast memory elements \cite{Lau2016} and neuromorphic chips \cite{Fukami2016}. Moreover, it has recently been suggested that current technology may have a lot to gain from antiferromagnet (AFM) materials. Indeed, manipulating AFM domains does not induce stray fields and has no fundamental speed limitations up to THz frequencies \cite{Jungwirth2016AFMreview}. Despite their ubiquitousness, AFM materials have, however, avoided much attention from technology due to an apparent lack of control over the AFM order parameter -- the N\'eel vector. Switching the N\'eel vector orientation by short electric pulses has been put forward only recently as the basis for AFM spintronics \cite{MacDonald2011,Gomonay2014,Zelezny2014}. The proposed phenomenon has been soon observed in non-centrosymmetric crystals such as CuMnAs \cite{Wadley2016, Fina2016, Zelezny2018, Saidl2017} and Mn$_2$Au \cite{Barthem2013, Jordan2015, Bhattacharjee2018}. It should be noted that in most cases AFMs are characterized by insulating type behavior \cite{Pandey2017}, limiting the range of their potential applications, e.g., for spin injection \cite{Tshitoyan2015}. Interestingly, antiferromagnetic Mn$_2$Au possesses a typical metal properties, inheriting strong spin-orbit coupling and high conductivity, and is characterized by collective modes excitations in THz range \cite{Bhattacharjee2018}.

Despite a lack of clarity concerning the microscopic mechanisms of the N\'eel vector switching, these experiments have been widely regarded as a breakthrough in the emerging field of THz spintronics \cite{Bhattacharjee2018, Gomonay2016AFM, Olejnik2018, Jungwirth2018, Wadley2016, Jungwirth2016AFMreview, Baltz2018, Jungwirth2018, Hoffman2018}. It has been suggested that current-induced N\'eel vector dynamics in an AFM is driven primarily by the so-called N\'eel spin-orbit torques \cite{Brataas2012, Hals2013, Zelezny2014, 2014MokrousovSOT, Ghosh2017, SmejkalAFM_2017, Zelezny2018, Zhou2018, Manchon2018, Moriyama2018, Li2019, Chen2019, Zhou2019, Zhou2019a, Bodnar2018}. The N\'eel spin-orbit torque originates in a non-equilibrium staggered polarization of conduction electrons on AFM sublattices \cite{Zelezny2014, SmejkalAFM_2017, Zelezny2018, Manchon2018}. Characteristic magnitude of the non-equilibrium staggered polarization and its relevance for the experiments with CuMnAs and Mn$_2$Au remain, however, debated. 

The N\'eel vector dynamics in an AFM is also strongly affected by an interplay between different types of Gilbert dampings. Unlike in a simple single-domain ferromagnet with a single sublattice, the Gilbert damping in an AFM is generally different on different sublattices and includes spin pumping from one sublattice to another. A proper understanding of Gilbert damping is of key importance for addressing not only the mechanism of spin pumping but also domain wall motion, magnon lifetime, AFM resonance width and many other related phenomena \cite{PhysRevMaterials.1.061401, Kamra2018, Mahfouzi2018a, Yuan_2019, Hals2011}. It is also worth noting that spin pumping between two thin ferromagnetic layers with antiparallel magnetic orientations share many similarities with Gilbert damping in a bipartite AFM \cite{Heinrich2003,Tserkovnyak2005}. 

A conduction electron mechanism for Gilbert damping in collinear ferromagnet requires some spin-orbit interaction to be present. It is, therefore, commonly assumed that spin-orbit interaction of electrons naturally enhances the Gilbert damping. Contrary to this intuition, we show that Rashba spin-orbit coupling does generally suppress one of the Gilbert damping coefficients and leads to the appearance of undamped non-equilibrium N\'eel vector precession modes in the AFM. 

\begin{figure}
\centering
\centerline{\includegraphics[width=\columnwidth]{{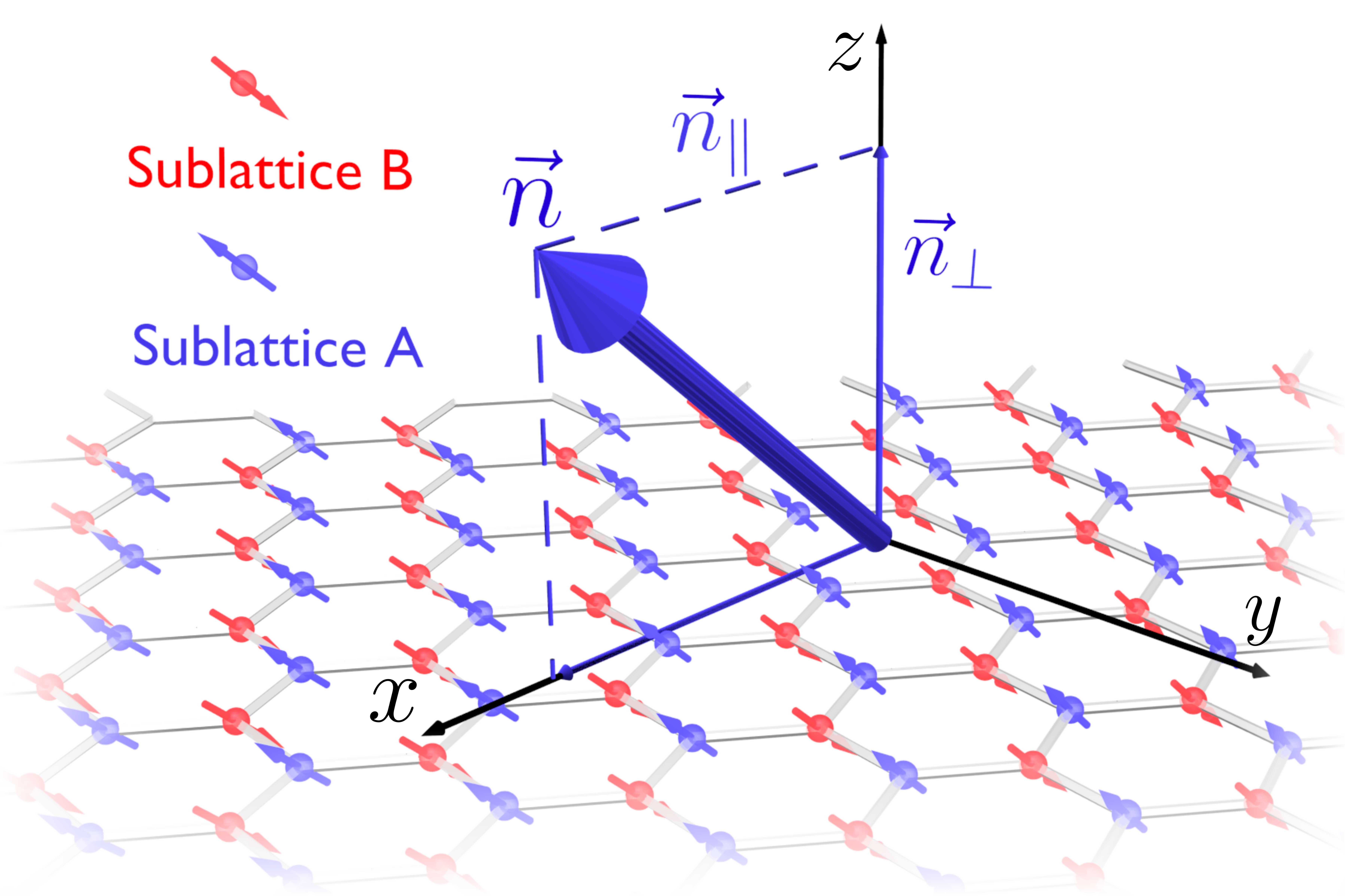}}}
\caption{A model of Rashba honeycomb antiferromagnet with two sublattices, $A$ and $B$, and on-site exchange interaction between localized momenta and conduction electrons (see Eq.~\ref{ex}). The large blue arrow represents the N\'eel vector vector, $\bm{n}$, that is in general, characterized by non-vanishing in-plane, $\bm{n}_\parallel$, and perpendicular-to-the-plane, $\bm{n}_\perp$, components. We refer to a specific coordinate system with $\hat{\bm{x}}$ axis chosen to be in the direction of $\bm{n}_\parallel$. 
}
\label{fig:lattice}
\end{figure}

Spin dynamics in a bipartite AFM is described in terms of two mutually orthogonal vector fields, namely the vector $\bb{n}(t)$ that is proportional to the N\'eel vector (difference between sublattice moments) and the vector $\bb{m}(t)$ that is proportional to the net magnetization (sum of sublattice moments) of a sample. Even though the AFM ground state corresponds to $\bb{m}=0$, it is widely understood that no N\'eel dynamics is possible without formation of a small but finite nonequilibrium magnetization $\bb{m}$. It appears, however, that Gilbert damping terms associated with the time dynamics of $\bb{m}(t)$ and $\bb{n}(t)$ are essentially different from a microscopic point of view.   

Indeed, the Gilbert damping that is proportional to $\partial_t\bb{n}$ is characterized by a coefficient $\alpha_n$, which is vanishing in the absence of spin-orbit interaction, much like it is the case in the ferromagnets. This behavior can be traced back to a spin-rotational symmetry of the collinear AFM. Indeed, the absolute value of $\bb{n}$ is conserved up to the order $m^2$. Thus, the dynamics of the N\'eel vector is essentially a rotation that does not change the conduction electron spectrum as far as the spin-rotation invariance is present. Breaking the spin-rotation symmetry by spin-orbit interaction induces, therefore, a finite $\alpha_n$, which is quadratic with respect to spin-orbit interaction strength.

In contrast, the Gilbert damping that is proportional to $\partial_t\bb{m}$ originates directly in the conduction electron scattering even in the absence of any spin-orbit interaction. The strength of the damping in a simple symmetric AFM is characterized by a coefficient $\alpha_m$, which is typically much larger than $\alpha_n$. As a rule, the spin-orbit interaction tends to suppress the coefficient $\alpha_m$ by restricting the ways in which electrons can damp their magnetic moments.  The condition $\alpha_m \gg \alpha_n$ has been indeed well documented in a metallic AFM \cite{PhysRevMaterials.1.061401, Mahfouzi2018a}. 

In this paper, we uncover the microscopic mechanism of strong and anisotropic Gilbert damping suppression due to the influence of spin-orbit interaction in a 2D AFM model on a honeycomb lattice. 

Below we focus mainly on the AFM in the regime of good metallic behavior, such that the Fermi energy of electrons exceeds by order of magnitude that of an effective $s$-$d$ exchange coupling between electron spins and localized AFM magnetic momenta.  In this case, the transition to the highly anisotropic regime takes place provided the characteristic spin-orbit energy $\lambda$ exceeds the scale $\hbar/\tau$, where $\tau$ is the electron scattering time. Alternatively, one may think of characteristic spin-orbit length becoming smaller than the mean free path of conduction electrons.  We show here that the splitting of 2D Fermi surfaces by spin-orbit interaction leads to a dramatic suppression of electron spin flips in certain directions. This results in a strong anisotropy of both Gilbert damping tensors $\hat{\alpha}_n$ and $\hat{\alpha}_m$, that get some of their principal components vanishing. This extreme anisotropy in the damping leads to essentially undamped N\'eel vector dynamics for certain nonequilibrium modes. 

In particular, we identify a specific undamped mode that corresponds to perpendicular-to-the-plane magnetization $\bb{m} \propto \hat{\bb{z}}$ and in-plane N\'eel vector $\bb{n}(t) \perp \hat{\bb{z}}$. The N\'eel vector corresponding to the mode has a precission around $\bb{m}$ with the frequency $J_\textrm{ex} m/\hbar$, where $J_\textrm{ex}$ is the value of the isotropic AFM exchange.

The presence of the undamped mode identified here, illustrates how lowering the symmetry of the electronic bath (by spin-orbit interaction) may induce a conservation law in the localized spin subsystem. Based on this microscopic mechanism we provide qualitative arguments in favor of a generality of the giant Gilbert damping anisotropy in a 2D metalic AFM with spin-orbit coupling. Even though the undamped mode cannot be associated with a single spin-wave or a magnon, its presence has a strong impact on the nonequilibrium N\'eel vector dynamics in 2D Rashba AFMs.

Apart from the Gilbert damping our results extend to cover conductivity and spin-orbit torques in the Rashba honeycomb AFM model. We also demonstrate how weak anisotropy of all these quantities emerge with Fermi energies approaching the AFM band gap. 

\section{Phenomenology of AFM dynamics}

In this paper, we choose to describe the AFM with a classical Heisenberg model for localized spins $\bb{S}^\textrm{X}= S \bb{n}^\textrm{X}$ on two sublattices $X=A,B$. The spins have the same modulus $S$ and antiparallel directions $\bb{n}^\textrm{A}=-\bb{n}^\textrm{B}$ in the ground state. The AFM Heisenberg model is coupled to an effective tight-binding model of conduction electrons (see Appendix~\ref{sec:appa}) by means of exchange interaction,
\be
\label{ex}
H_\mathrm{sd}=-J \s_{i} \s_{\sigma\sigma'}\bb{S}_i\cdot \bb{\sigma}_{\sigma\sigma'}c^\dagger_{i\sigma}c\0_{i\sigma'},
\e
where $J$ stands for an $s$-$d$-like exchange energy that is the same on $A$ and $B$ sublattices, the operators $c\h_{i\sigma}$ ($c\0_{i\sigma}$) are the standard creation (annihilation) operators for an electron on the lattice site $i$ with the spin index $\sigma$, and the notation $\bb{\sigma}=(\sigma_x,\sigma_y,\sigma_z)$ represents the three-dimensional vector of Pauli matrices. 

The real-time dynamics of AFM is, then, defined by two coupled differential equations (Landau-Lifshitz-Gilbert equations) on the unit vectors $\bb{n}^\textrm{A}$ and $\bb{n}^\textrm{B}$, 
\beml
\label{basicEQ}
\begin{align}
\dot{\bb{n}}^\textrm{A} &= \bb{H}^\textrm{A}\times\bb{n}^\textrm{A}  + (J\mathcal{A}/\hbar)\,\bb{n}^\textrm{A}\times \bb{s}^\textrm{A},\\
\dot{\bb{n}}^\textrm{B}&= \bb{H}^\textrm{B}\times\bb{n}^\textrm{B} +(J\mathcal{A}/\hbar)\,\bb{n}^\textrm{B}\times \bb{s}^\textrm{B},
\end{align}
\eml
where dot stands for the time derivative, $\bb{s}^\textrm{X}$ is the spin density of conduction electrons on the sublattice $X$,
\be
\bb{s}^\textrm{A,B}(\bb{r})= \frac{1}{2} \s_{i\sigma\sigma'} \lt\la c\h_{i\sigma}\bb{\sigma}_{\sigma\sigma'} c\0_{i\sigma'} \rt\ra\;\frac{2}{\mathcal{A}},
\e
and $\mathcal{A}$ is the area of the unit cell in the AFM. The notations $\bb{H}^\textrm{A,B}$ refer to effective fields on the sublattices $A$ and $B$ that are defined by the Heisenberg model. 

For an isotropic antiferromagnet, one finds an effective field \cite{Gomonay2014} $\bb{H}^\textrm{A}+\bb{H}^\textrm{B}= J_\textrm{ex}\bb{m}/\hbar+2\bb{H}$, where $\bb{H}$ is an external magnetic field in frequency units and $J_\textrm{ex}$ is a direct antiferromagnetic exchange energy that is one of the largest energies in the problem. In turn, the combination $\bb{H}^\textrm{A}-\bb{H}^\textrm{B}$ is proportional to magnetic anisotropy that we do not specify in this paper.  

Magnetization dynamics in AFM is conveniently formulated in terms of the N\'eel and magnetization vectors,
\be
\bb{n}=\lt(\bb{n}^\textrm{A}-\bb{n}^\textrm{B}\rt)/2,\qquad \bb{m}= \lt(\bb{n}^\textrm{A}+\bb{n}^\textrm{B}\rt)/2,
\e
that remain mutually perpendicular $\bb{n}\cdot \bb{m}=0$ and yield the constraint $n^2+m^2=1$. The dynamics necessarily induces a finite nonequilibrium magnetization vector $\bb{m}$, while the condition $m\ll 1$ remains to be fulfilled.  

From Eqs.~(\ref{basicEQ}) we obtain
\beml
\label{AFMEOM}
\begin{align}
\label{ndot}
\dot{\bb{n}} &= -\Omega\, \bb{n}\times\bb{m} +\bb{H}\times\bb{n}+\bb{n}\times\bb{s}^++\bb{m}\times\bb{s}^-,\\
\label{mdot}
\dot{\bb{m}} &= \bb{H}\times\bb{m}+\bb{m}\times\bb{s}^++\bb{n}\times\bb{s}^-,
\end{align}
\eml
where $\Omega=2J_\textrm{ex}S/\hbar$ and $\bb{s}^{\pm}= J \mathcal{A}\,(\bb{s}^\textrm{A}\pm\bb{s}^\textrm{B})/2\hbar$. In Eqs.~(\ref{AFMEOM}) we have deliberately skipped terms that are induced by anisotropy of AFM exchange since the latter depend on particularities of the AFM Heisenberg model that we do not discuss here.  

The vector $\bb{s}^+$ is proportional to average polarization of conduction electrons, while the vector $\bb{s}^-$ is proportional to the staggered polarization. The quantities $\bb{s}^\pm=\bb{s}_0^\pm+\delta\bb{s}^\pm$ contain equilibrium contributions $\bb{s}_0^\pm$ that characterize various interactions induced by conduction electrons. These contributions do renormalize the parameters of the AFM Heisenberg model and are not the subject of the present paper. 

\begin{figure}
\centering
\centerline{\includegraphics[width=\columnwidth]{{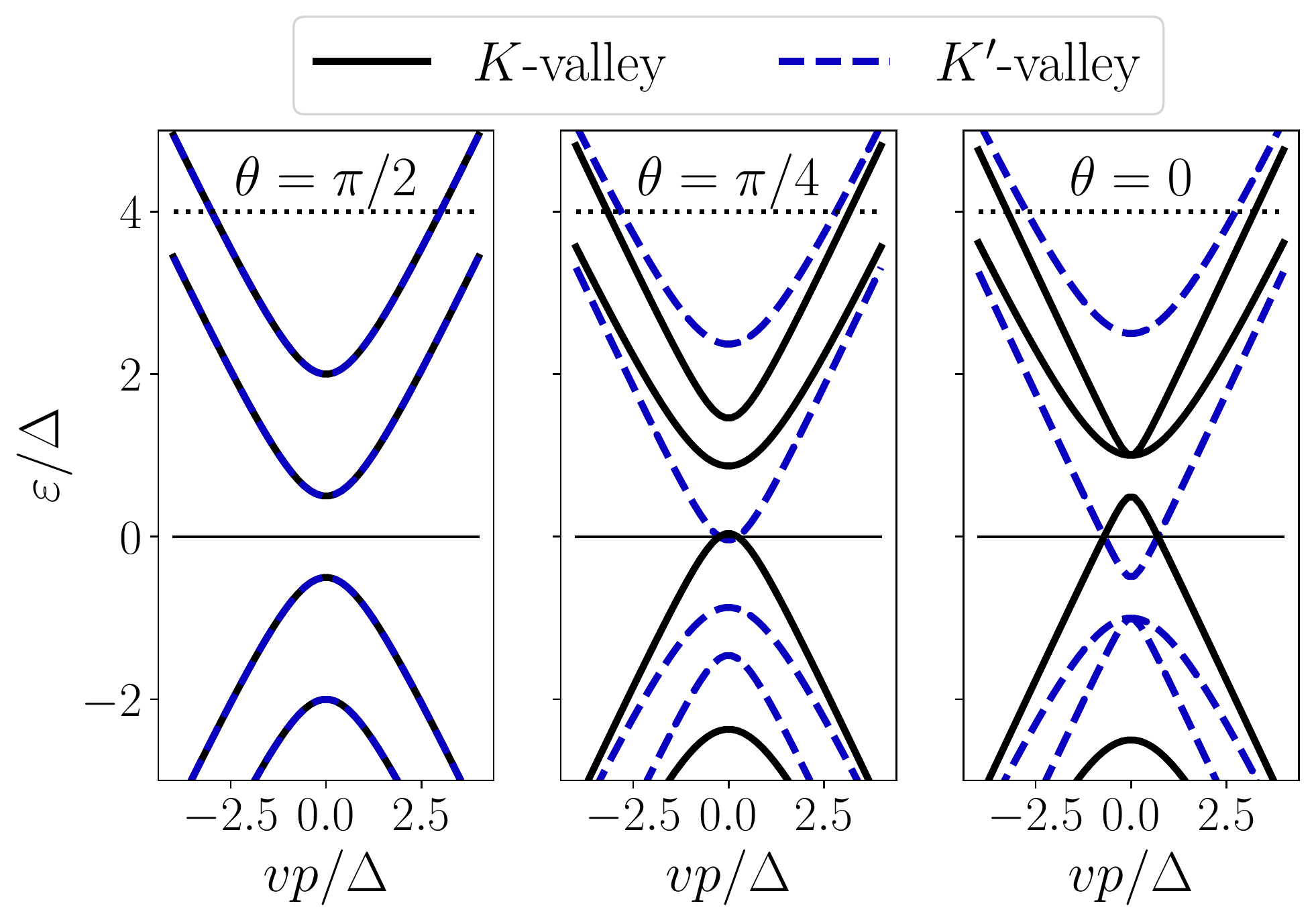}}}
\caption{Electronic band structure of the honeycomb AFM model of Eq.~(\ref{eff}) for different orientations of the N\'eel vector ($n_z=\cos\theta$). Two-dimensional momenta $\bb{p}$ are measured with respect to the wave-vectors $\bm{K}$ and $\bm{K}^\prime$ that specify two nonequivalent valleys. Deviation of the N\'eel vector from the perpendicular-to-the plane configuration ($\theta=0$) lifts the valley degeneracy. We restrict our analysis to the metallic regime with Fermi energies corresponding to two Fermi surfaces per valley (an example is shown by a black dotted line). The energy scale $\Delta$ characterizes the strength of $s$-$d$ exchange interaction.
}
\label{fig:spectrum}
\end{figure}

The nonequilibrium contributions $\delta\bb{s}^{\pm}$ originate from various forces applied to conduction electrons. One natural example is the electric field that not only induces an electric current in the sample but also contributes to $\delta\bb{s}^{\pm}$. The electric field can be further related to electric current by the resistivity tensor. The response of spin densities to electric current defines the so-called spin-orbit torques in Eqs.~(\ref{AFMEOM}) that we also compute.

Similarly, the response of $\delta\bb{s}^{\pm}$ to the time derivatives $\dot{\bb{n}}$ and $\dot{\bb{m}}$ describe various types of Gilbert damping induced by conduction electrons. Quite generally, such a response can be written in the form of a tensor
\be
\label{response}
\bpm 
\delta \bb{s}^{+}\\
\delta \bb{s}^{-}
\epm=
\bpm 
\hat{\alpha}_m & \hat{\alpha}_{mn}\\
\hat{\alpha}_{nm} & \hat{\alpha}_n
\epm
\bpm 
\dot{\bb{m}}\\
\dot{\bb{n}}
\epm,
\e
where all tensor components may themselves depend on the vectors $\bb{n}$ and $\bb{m}$. 

Gilbert dampings, in their original meaning, correspond to the contributions to $\delta\bb{s}^{\pm}$ that are symmetric under the time reversion. 
The terms that change sign should, more appropriately, be referred to as effective spin renormalizations. Both types of terms are, however, obtained from the microscopic analysis of the Gilbert damping tensors in Eq.~(\ref{response}) similarly to the case of ferromagnets \cite{AdoSTTGD2019}.

Time reversion, mentioned above, applies exclusively to the Heisenberg model, while keeping the tight-binding model (a bath) non-reversed. In other words we do not reverse the electron scattering time $\tau$. Such a definition helps to identify the dissipative (even with respect to the time reversion) contributions to $\delta\bb{s}^{\pm}$ that describe Gilbert dampings. These contributions must, however, change sign under the transformation $\tau \to -\tau$, because spin densities $\bb{s}^\pm$ are always odd with respect to complete time reversion (the one which also includes that of the electron bath). We will see below, indeed, that all Gilbert dampings are proportional to the scattering time $\tau$ in the same way as the longitudinal conductivity does.  

Before we proceed with the microscopic analysis of $\delta\bb{s}^{\pm}$ for a particular model, it is instructive to draw some general consequences for Eqs.~(\ref{AFMEOM}) based on symmetry arguments in the case of collinear AFM with sublattice symmetry and spin-rotational invariance (i.\,e. for vanishing spin-orbit interaction).  

Assuming that deviations from the AFM ground state remain small we shall limit ourselves to the linear order in $\bb{m}$ in Eq.~(\ref{even}) and to the quadratic order in $\bb{m}$ in Eq.~(\ref{staggered}).  Thus, we shall retain terms up to linear order in $\bb{m}$ in the tensors $\hat{\alpha}_{m}$, $\hat{\alpha}_{nm}$, and $\hat{\alpha}_{mn}$ and terms up to quadratic order in $\bb{m}$ in $\hat{\alpha}_n$. 

Mixing tensors $\hat{\alpha}_{mn}$ and $\hat{\alpha}_{nm}$ must be odd in $\bb{m}$, which implies, for our precision, a linear in $\bb{m}$ approximation. As a result, the sublattice symmetry (the symmetry with respect to renaming $A$ and $B$) prescribes that the mixing tensors must also be linear in $\bb{n}$. In the absence of spin-orbit coupling we are also restricted by spin-rotation invariance that (together with the sublattice and time-reversion symmetries) dictates the following form of the Gilbert damping contributions to the non-equilibrium spin densities
\beml
\label{gen}
\begin{align}
\label{even}
\delta \bb{s}^{+} & = \alpha_m \dot{\bb{m}}  \!+\! \alpha'_m \bb{n}  \!\times\! (\bb{n} \!\times\! \dot{\bb{m}}) \!+\!\alpha_{mn}\bb{m}  \!\times\!  (\bb{n}  \!\times\!  \dot{\bb{n}}),\\
\delta \bb{s}^{-} &= \alpha_n \dot{\bb{n}} \!+\! \alpha'_n \bb{m}  \!\times\! (\bb{m} \!\times\! \dot{\bb{n}}) + \alpha_{nm} \bb{n}\!\times\!(\bb{m}\!\times\! \dot{\bb{m}}),
\label{staggered}
\end{align}
\eml
where all coefficients are assumed to be constants.  

It is easy to see that the vector forms $\bb{n}\times(\bb{m}\times\dot{\bb{n}})$ and $\bb{m}\times(\bb{n}\times \dot{\bb{m}})$, which could have respectively entered the spin densities $\delta\bb{s}^+$ and $\delta\bb{s}^-$, do not contribute to Eqs.~(\ref{AFMEOM}) in the precision explained above.  
Substitution of Eqs.~(\ref{gen}) into Eqs.~(\ref{AFMEOM}) gives 
\beml
\label{AFMEOM2}
\begin{align}
\label{ndot2}
&\dot{\bb{n}} = -\Omega\, \bb{n}\!\times\!\bb{m} +\bb{H}\!\times\!\bb{n}+\bar{\alpha}_m\,\bb{n}\!\times\!\dot{\bb{m}}+\alpha_n\,\bb{m}\!\times\! \dot{\bb{n}},\\
\label{mdot2}
&\dot{\bb{m}} =\bb{H}\times\bb{m}+\alpha_n\,\bb{n} \times \dot{\bb{n}}\n\\
&\;+\bar{\alpha}_m \bb{m} \times \dot{\bb{m}} +\gamma (\bb{n}\times\bb{m})(\bb{n}\cdot\dot{\bb{m}}) - \alpha_n'm^2\bb{n}\times\dot{\bb{n}} ,
\end{align}
\eml
where $\bar{\alpha}_m=\alpha_m\!-\!\alpha'_m$ and $\gamma=\alpha_{mn}\!+\!\alpha_{nm}\!+\!\alpha'_m\!-\!\alpha'_n$. Discarding the three last terms in Eq.~(\ref{mdot2}), which are all of the second order in $\bb{m}$, we indeed arrive at a set of Gilbert damping terms that is  widely used in the AFM literature \cite{Kamra2018,PhysRevMaterials.1.061401,Yuan_2019}. 

The symmetry consideration behind Eqs.~(\ref{AFMEOM2}) has essentially relied upon the spin-rotation invariance. This also implies $\alpha_n=0$ as has been pointed out in the introductory section. The coefficient $\alpha_m$ can, in turn, be finite and large, even in the absence of spin-orbit interaction. As we will show below, the presence of spin-orbit interaction does not only provide us with a finite $\alpha_n$ but also drastically change the symmetry structure of Eqs.~(\ref{AFMEOM2}). We will demonstrate that the onset of spin-orbit interaction strongly affects the coupling of the localized spin subsystem to the electron bath (described by the tight-binding model) resulting in a strong reduction in the ability of conduction electrons to flip spins in certain directions and, therefore, to impose a friction on magnetization dynamics. 

In the following, we turn to the microscopic analysis of the conductivity (Sec.~\ref{sec:cond}), spin-orbit torques (Sec.~\ref{sec:sot}) and Gilbert dampings (Sec.~\ref{sec:gd}) in a particular model of Rashba honeycomb AFM that has been put forward recently by some of the authors \cite{Sumit2019}. Rashba spin-orbit interaction breaks spin-rotational invariance of the model by singling out the direction $\hat{\bb{z}}$ perpendicular to the 2D plane. We, therefore, investigate how such spin-rotation breaking manifests itself in the anisotropy of the abovementioned quantities.  

\section{Microscopic model}

For the sake of a microscopic analysis we adopt a sublattice symmetric $s$-$d$-like model of a 2D honeycomb antiferromagnet with Rashba spin-orbit coupling, that was introduced in Ref.~\onlinecite{Sumit2019}. The energy dispersion of this model is illustrated schematically in Fig.~\ref{fig:spectrum}.  
The low energy model for conduction electrons responsible for the dispersion in Fig.~\ref{fig:spectrum}, is described by an effective Hamiltonian (see Appendix ~\ref{sec:appa}) that in a valley-symmetric representation reads
\be
\label{eff}
H^\textrm{eff}= v\, \bb{p}\cdot\bb{\Sigma}+\tfrac{1}{2}\lambda\lt[\bb{\sigma}\times\bb{\Sigma}\rt]_{\hat{z}} - \Delta\,\bb{n}\cdot\bb{\sigma}\,\Sigma_z\Lambda_z + V(\bb{r}).
\e
Here $\bb{\Sigma}$, $\bb{\Lambda}$, and $\bb{\sigma}$ are the vectors of Pauli matrices in sublattice, valley and spin space, respectively, $v$ is the characteristic Fermi velocity, while $\lambda$ and $\Delta=JS$ are the energy scales characterizing the strength of Rashba spin-orbit coupling and $s$-$d$-like exchange energy, correspondingly. 

The term $V(\bb{r})$ stands for a scalar Gaussian white-noise disorder potential, which is proportional to the unit matrix in sublattice, valley and spin space. The potential has a zero mean value $\la V(\bb{r}) \ra=0$ and is fully characterized by the pair correlator,
\be
\lt\la V(\bb{r}) V(\bb{r}') \rt\ra = 2\pi (\hbar v)^2 \alpha_\textrm{d}\;\delta(\bb{r}-\bb{r}'),
\e
where the angular brackets denote the averaging over disorder realizations. The dimensionless parameter $\alpha_\textrm{d}\ll 1$ quantifies the disorder strength. 

The disorder potential is responsible for a momentum relaxation of conduction electrons. Exchange interaction and spin-orbit scattering (or the scattering on a non-collinear configurations with $\bb{m}\neq 0$) enable coupling between localized angular momenta and kinetic momenta of electrons. Together these mechanisms form a channel to dissipate angular momentum of localized spins into the lattice. Thus, our model provides us with a microscopic framework to study dissipative quantities such as Gilbert dampings, anti-damping spin-orbit torques and conductivity that we compute below. We also note that the computation of spin-relaxation time can be directly related to our analysis of Gilbert damping \cite{Hankiewicz2007,Manchon2017}. 

The spectrum of the model (\ref{eff}) with $V(\bb{r})=0$ consists of two electron and two hole branches for each of the valleys as illustrated in Fig.~\ref{fig:spectrum},
\beml
\label{spectrum}
\begin{align}
\label{spectrume}
\epsilon^e_{\pm,\varsigma}(p)&=\sqrt{v^2p^2+\Delta^2\pm \varsigma\lambda\Delta n_z+\lambda^2/4} \mp \lambda/2,\\
\epsilon^h_{\pm,\varsigma}(p)&=-\sqrt{v^2p^2+\Delta^2\mp \varsigma\lambda\Delta n_z+\lambda^2/4} \pm \lambda/2,
\end{align}
\eml
where $\varsigma=\pm$ is the valley index. All spectral branches are manifestly isotropic with respect to the direction of the electron momentum $\bb{p}$ irrespective of the N\'eel vector orientation (as far as $\bb{m}=0$). 

In order to limit the complexity of our microscopic analysis we restrict ourselves to the metallic regime that corresponds to the Fermi energy $\ep_F > \Delta+\lambda$ above the minimum of the top electron branches, $\epsilon^e_{+,\varsigma}(p)$, as shown schematically in Fig.~\ref{fig:spectrum}. Note that the Fermi energy $\ep_F$ is counted in the model from the center of the AFM gap. 
We also focus on the limit of weak disorder $\ep_F \tau/\hbar\gg 1$ where $\tau = \hbar/(\pi\alpha_\textrm{d}\ep_F)$ stands for the electron scattering time. Also, in order to describe spin-orbit induced anisotropy we find it convenient to decompose the N\'eel vector (as well as the magnetization vector) to the in-plane and perpendicular-to-the-plane components as $\bb{n}= \bb{n}_\parallel+\bb{n}_\perp$, where $\bb{n}_\perp=n_z\hat{\bb{z}}$. 

\section{Conductivity}\label{sec:cond}
The electric conductivity in the metallic regime is dominated by electron diffusion. Despite the fact that the Fermi surface (line) is isotropic and does not depend on the direction of $\bb{n}_\parallel$, the conductivity appears to be weakly anisotropic with respect to in-plane rotations of the N\'eel vector due to the onset of spin-orbit interaction. 
In particular, for $n_z=0$ we find the diagonal conductivity components
\beml
\label{conductivity}
\begin{align}
\label{eq:long_cond}
&\sigma_{xx}= \frac{4e^2}{h}\,\frac{\ep_F \tau}{\hbar}\,\frac{\ep_F^2-\Delta^2}{\ep_F^2+3\Delta^2},\\
&\sigma_{yy}= \sigma_{xx}+\frac{4e^2}{h}\,\frac{\ep_F \tau}{\hbar}\, \frac{\ep_F^2}{\ep_F^2+\Delta^2}
\frac{\lambda^2\Delta^2}{\ep_F^4+\ep_F^2\Delta^2+2\Delta^4},
\end{align}
\eml
where the principal axes correspond to choosing $\hat{\bm{x}}$ direction along $\bb{n}_\parallel$ (see Fig.~\ref{fig:lattice}). In the deep metal regime, and for a general direction of $\bb{n}$, this anisotropy is evidently small
\be
\label{ani}
\frac{\rho_{xx}-\rho_{yy}}{\rho_{xx}+\rho_{yy}}=\frac{\lambda^2\Delta^2}{\ep_F^4}(1-n_z^2), \quad \ep_F\gg \lambda+\Delta,
\e
where $\rho_{aa}=1/\sigma_{aa}$ is the corresponding resistivity tensor component. We note that the anomalous Hall conductivity is identically vanishing in the model of Eq.~(\ref{eff}). 

The results of Eq.~(\ref{conductivity}) and all subsequent results of our paper are technically obtained from linear response Kubo formulas evaluated in the diffusive approximation (ladder diagram summation). The details of these calculations can be found in Appendixes~\ref{sec:appb}, \ref{sec:appc}, and \ref{sec:appd}. 

\section{Spin-orbit torque}\label{sec:sot}

Before proceeding with the microscopic analysis of Gilbert damping we shall discuss the effects of spin-orbit induced anisotropy for spin-orbit torques in the model of Eq.~(\ref{eff}). Since this anisotropy appears to be weak in the metal regime, we shall touch on it only briefly. 

As was already mentioned, the spin-orbit torques originate in the response of nonequilibrium spin polarizations $\delta\bb{s}^\pm$ to electric current. Technically, we compute first the response of $\delta\bb{s}^\pm$ to electric field and, then, express the electric field in terms of 2D electric current density $\bb{j}$ using the conductivity tensor of Eq.~(\ref{conductivity}).

\begin{figure}
\centering
\centerline{\includegraphics[width=\columnwidth]{{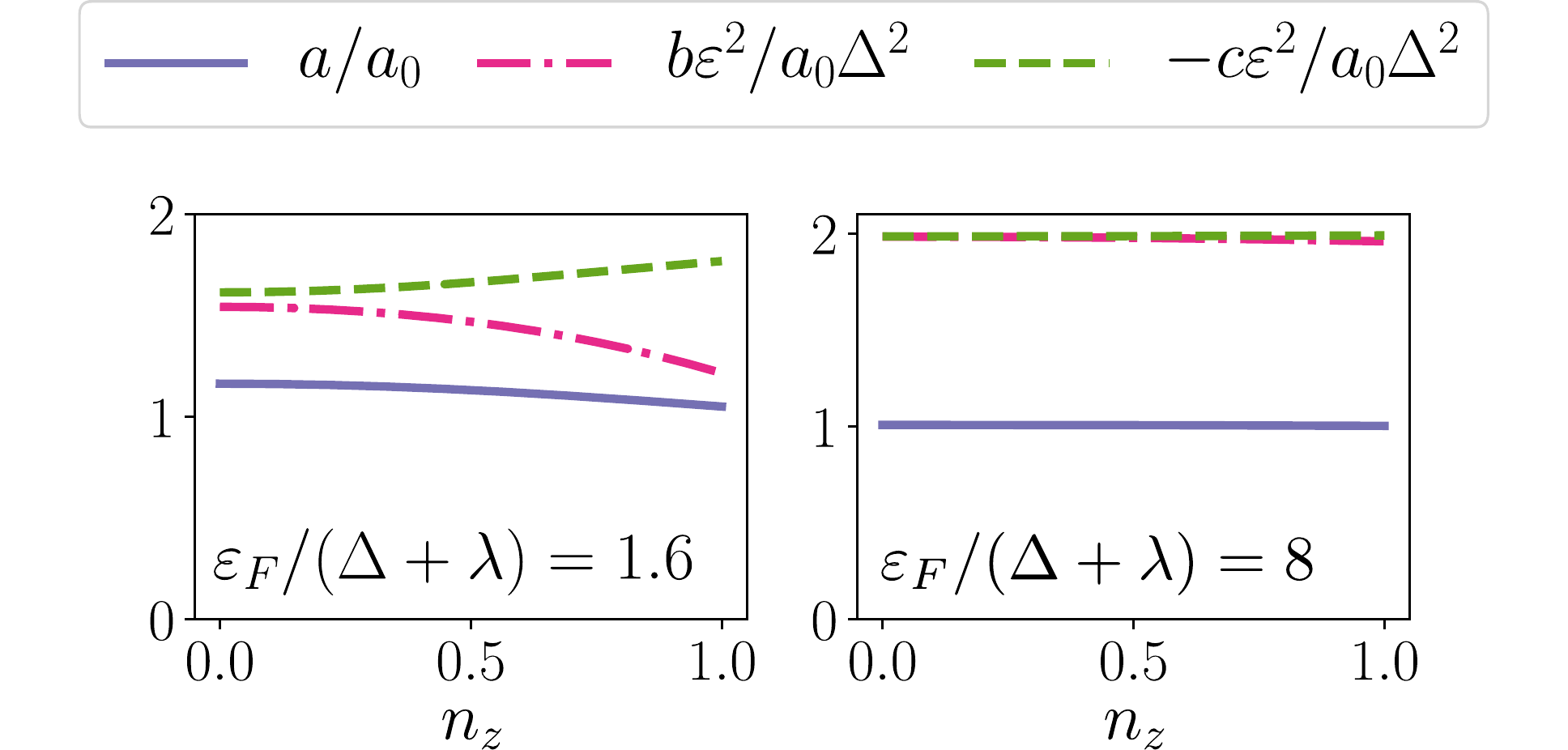}}}
\caption{The coefficients $a$, $b$, and $c$ in Eq.~(\ref{sp}) as a function of the direction of the N\'eel vector, $n_z=\cos\theta$, for two different Fermi energies: $\ep_F=4\Delta$ (left panel) and  $\ep_F=16\Delta$ (right panel). We use $\lambda=1.5\Delta$.  For $n_z=0$ the results correspond to Eqs.~(\ref{SOT}). 
}
\label{fig:abc}
\end{figure}

A straightforward computation of such a response gives $\delta\bb{s}^-=0$ (see Appendixes~\ref{sec:appb} and \ref{sec:appc} for more detail) and 
\begin{align}
\label{sp}
\delta\bb{s}^+= \,&a(n_z^2)\;\hat{\bb{z}}\times \bb{j} + b(n_z^2)\, \bb{n}_\parallel\times(\bb{n}_\parallel\times \lt(\hat{\bb{z}}\times \bb{j}\rt)) \n\\
&+ c(n_z^2)\, \bb{n}_\parallel \times(\bb{n}_\perp\times \lt(\hat{\bb{z}}\times \bb{j}\rt)),
\end{align}
where the coefficients $a$, $b$ and $c$ do generally depend on $n_z^2=1-n_x^2-n_y^2$ and are shown in Fig.~\ref{fig:abc}. 
It is appropriate to recall here that the computation of the responses from the model of Eq.~(\ref{eff}) refers to the case when $\bb{m}=0$. The symmetry form of Eq.~(\ref{sp}) in this case has been also established recently from numerical simulations \cite{Sumit2019}.

Importantly, the first term in the right-hand side of Eq.~(\ref{sp}) represents the well-known Rashba-Edelstein effect \cite{Edelstein1990}, while the other two terms represent higher harmonics of the same field-like effect that arise due to spin-rotation symmetry breaking. Anti-damping like torques (that are even under time-reversal) are vanishing in the model due to the valley symmetry constraint. This symmetry reads $\Lambda_x H[\bb{n}] \Lambda_x = H[-\bb{n}]$, from which it follows that the response of $\delta\bb{s}^+$ to charge current must be an even function of $\bb{n}$. 

The behavior of the coefficients $a$, $b$ and $c$ as a function of $n_z$ is illustrated in Fig.~\ref{fig:abc} for two different choices of the Fermi energy. For in-plane N\'eel vector orientations ($n_z=0$) we find 
\beml
\label{SOT}
\begin{align}
&a=a_0 \frac{1+3\delta^2}{1+2\bar{\lambda}^2\delta^2+\delta^4-2\delta^6},\\
&b=2\, a\,\delta^2\,\frac{1-2\bar{\lambda}^2-4\delta^2+\delta^4}{1+2\delta^2-3\delta^4},\\
&c=-2\, a\,\delta^2\,\frac{1+2\bar{\lambda}^2\delta^2-2\delta^2-3\delta^4-4\delta^6}{1+4\delta^2+5\delta^4+6\delta^6},
\end{align}
\eml
where
\be
a_0=\frac{\mathcal{A}J}{e\hbar v}\,\frac{\lambda}{\ep_F}, \qquad \bar{\lambda}=\frac{\lambda}{\ep_F},\qquad\delta=\frac{\Delta}{\ep_F}.
\e
In the metal regime, $\ep_F\gg \lambda+\Delta$, the results of Eqs.~(\ref{SOT}) are reduced to 
\be
\label{SOTlargeE}
a=\frac{\mathcal{A}J}{e\hbar v}\,\frac{\lambda}{\ep_F}, \qquad b=-c=2\,\frac{\mathcal{A}J}{e\hbar v}\,\frac{\lambda}{\ep_F}
\lt(\frac{\Delta}{\ep_F}\rt)^2.
\e
One can, therefore, see that the high harmonics terms (proportional to $b$ and $c$) become irrelevant in the metal regime.  

Vanishing response of the staggered polarization, $\delta\bb{s}^-=0$, for the model of Eq.~(\ref{eff}) is a simple consequence of the sublattice symmetry. As shown below the presence of a finite, though small, $\bb{m}$ breaks such a symmetry and leads to a finite $\delta\bb{s}^-$. Taking into account a linear in $\bb{m}$ term in the Hamiltonian is also necessary to obtain finite mixed Gilbert damping tensors $\hat{\alpha}_{nm}$ and $\hat{\alpha}_{mn}$ in Eq.~(\ref{response}).   

A low-energy model that takes into account finite magnetization vector reads (see also Appendix~\ref{sec:appd})
\be
\label{full}
H=H^\textrm{eff}-\Delta\,\bb{m}\cdot\bb{\sigma},
\e
where $H^\textrm{eff}$ is given by Eq.~(\ref{eff}). The conductivity tensor does not acquire a linear in $\bb{m}$ terms in the leading order with respect to the large metal parameter $\ep_F\tau/\hbar$, because the anomalous Hall effect remains subleading with respet to the metal parameter. Similarly, the result of Eq.~(\ref{sp}) is not affected by the linear in $\bb{m}$ corrections. 

However, the direct computation of the staggered polarization response (in the linear order with respect to $\bb{m}$) gives rise to a finite result. In the limit of large Fermi energy $\ep_F\gg \lambda+\Delta$, we find 
\begin{align}
&\delta\bb{s}^-= \frac{\mathcal{A}J}{e\hbar v}\,\frac{\lambda}{\ep_F}
\lt(\frac{\Delta}{\ep_F}\rt)^2\,
\Big[2\,\bb{n}_{\perp} \!\times\!(\bb{m}_{\perp} \!\times\! (\hat{\bb{z}}\!\times\!\bb{j}))
\\
&+2\,\bb{m}_{\parallel} \!\times\!(\bb{n}_{\perp} \!\times\! (\hat{\bb{z}}\!\times\!\bb{j})) -3\,\bb{n}_{\parallel}\!\times\! (\bb{m}_{\perp}\! \times\! (\hat{\bb{z}}\!\times\!\bb{j}))\n \Big],
\label{stagg}
\end{align}
where the overall strength of the effect is again of the order of the coefficients $b$ and $c$. This makes the effects of nonequilibrium staggered polarization (including the celebrated N\'eel spin-orbit torque) irrelevant in the metal regime. Indeed, staggered polarization can hardly be induced by electrons with wavelengths that strongly exceed the distance between $A$ and $B$ sublattices.  

The results of Eqs.~(\ref{sp}), (\ref{SOT}) clearly suggest that the only torques surviving in the large energy limit are those related to non-equilibrium polarization $\delta\bb{s}_+=a_0 \,\hat{\bb{z}}\times\bb{j}$, which is nothing but the standard Rashba-Edelstein effect \cite{Edelstein1990}. These torques have a form $\bb{T}_n= a_0 \,\bb{n}\times(\hat{\bb{z}}\times\bb{j})$ in the right-hand side of Eq.~(\ref{ndot}) and $\bb{T}_m= a_0 \,\bb{m}\times(\hat{\bb{z}}\times\bb{j})$ in the right-hand side of Eq.~(\ref{mdot}). The anisotropy of torques is, however, irrelevant in this limit.  

\section{Gilbert damping}\label{sec:gd}

Surprisingly, the situation is different when we consider Gilbert damping terms. In this case we find that the giant anisotropy of Gilbert damping persists to arbitrarily large Fermi energy as soon as spin orbit energy $\lambda$ exceeds $\hbar/\tau$. The latter condition ensures that the scattering between spin-split subbands is suppressed. 

The direct computation of the Gilbert damping tensors for $\lambda\gg \hbar/\tau$ gives
\beml
\label{gilbert}
\begin{align}
&\delta\bb{s}^+= \alpha_m^\parallel\,\dot{\bb{m}}_\parallel + \gamma \bb{\Gamma}_{mm}+ \gamma\bb{\Gamma}_{mn},\\
&\delta\bb{s}^-= \alpha_n^\perp\,\dot{\bb{n}}_\perp +\gamma \bb{\Gamma}_{nm}+\gamma \bb{\Gamma}_{nn},
\end{align}
\eml
where the terms $\Gamma_{ab}$ contain various vector forms. 

Far in the metal regime, $\ep_F\gg \lambda+\Delta$, we find
\beml
\label{coeffGD}
\begin{align}
&\alpha_m^\parallel= 2\,\frac{\ep_F\tau}{\hbar}\;\frac{\mathcal{A} J^2 S}{\pi\hbar^2v^2}\; \lt[1-\frac{\Delta^2}{\ep_F^2}(2+n^2_z) + \dots\rt],\\
&\alpha_n^\perp = \frac{\ep_F\tau}{\hbar}\;\frac{\mathcal{A} J^2 S}{\pi\hbar^2v^2} \,\lt[\lt(\frac{\lambda}{\ep_F}\rt)^2+\dots\rt],\\
&\gamma= 2\,\frac{\ep_F\tau}{\hbar}\;\frac{\mathcal{A} J^2 S}{\pi\hbar^2v^2}\,\lt[\lt(\frac{\Delta}{\ep_\textrm{F}}\rt)^2+\dots\rt],
\end{align}
\eml
while the vectors forms $\Gamma_{ab}$ can be written as
\beml
\begin{align}
\bb{\Gamma}_{mm}=&\,
\bb{n}\!\times\!(\bb{n}\!\times\!\dot{\bb{m}}) + \bb{n}_\parallel\!\times\!(\bb{n}_\parallel\!\times\!\dot{\bb{m}}_\parallel) \n\\ 
&- 2 \bb{n}_\parallel\!\times\!(\bb{n}_\parallel\!\times\!\dot{\bb{m}}_\perp),\\
\bb{\Gamma}_{mn}=&\,
\bb{n}\!\times\!(\bb{m}_\parallel\!\times\!\dot{\bb{n}}_\perp) 
- \bb{m}_\perp\!\times\!(\bb{n}_\parallel\!\times\!\dot{\bb{n}}_\perp)\n\\
&+\bb{n}_\perp\!\times\!(\bb{m}_\perp\!\times\!\dot{\bb{n}}_\parallel) 
- \bb{n}_\parallel\!\times\!(\bb{m}_\parallel\!\times\!\dot{\bb{n}}_\parallel)\n\\
&- 3\bb{m}_\perp\!\times\!(\bb{n}_\perp\!\times\!\dot{\bb{n}}_\parallel),\\
\bb{\Gamma}_{nm}=&\,2\bb{n}_\parallel\!\times\!(\bb{m}_\parallel\!\times\!\dot{\bb{m}}_\perp)+ 
2\bb{m}_\parallel\!\times\!(\bb{n}_\perp\!\times\!\dot{\bb{m}}_\parallel)\n\\
&-\bb{n}_\perp\!\times\!(\bb{m}_\perp\!\times\!\dot{\bb{m}})+2 \bb{m}_\perp\!\times\!(\bb{n}\!\times\!\dot{\bb{m}})\n\\
&+ \bb{m}_\parallel\!\times\!(\bb{n}\!\times\!\dot{\bb{m}}_\perp) 
- \bb{m}_\perp\!\times\!(\bb{n}_\perp\!\times\!\dot{\bb{m}}_\parallel),\\
\bb{\Gamma}_{nn}=&\,-\bb{m}\times(\bb{n}\times\dot{\bb{n}}_\parallel).
\end{align}
\eml
Thus, we see from Eqs.~(\ref{coeffGD}) that the coefficients $\alpha_n^{\perp}$ and $\gamma$ are vanishingly small in the metal regime. Moreover, in the limit $\ep_F\gg \Delta$ the only non-vanishing contributions to Gilbert dampings are given by the first terms on the right-hand sides of Eqs.~(\ref{gilbert}) that are manifestly anisotropic.


The onset of spin-orbit interactions therefore makes Gilbert dampings ultimately anisotropic, also in the deep metal regime. This is in contrast to conductivity and spin-orbit torques that are quickly becoming isotropic in the metal limit. 
For $\ep_F\gg \lambda+\Delta$, we find the well known Landau-Lifshitz-Gilbert equations
\begin{align}
&\dot{\bb{n}} = -\Omega\, \bb{n}\!\times\!\bb{m} +\bb{H}\!\times\!\bb{n}+\bar{\alpha}^\parallel_m\,\bb{n}\!\times\!\dot{\bb{m}}_\parallel+\alpha_n^\perp\,\bb{m}\!\times\! \dot{\bb{n}}_\perp,\n\\
&\dot{\bb{m}} =\bb{H}\times\bb{m}+\alpha_n^\perp\,\bb{n} \times \dot{\bb{n}}_\perp
+\bar{\alpha}_m^\parallel \bb{m} \times \dot{\bb{m}}_\parallel,
\label{AFMEOM3}
\end{align}
where we again omit terms that originate e.\,g. from magnetic anisotropy of the AFM. 
Eqs. (\ref{AFMEOM3}) are clearly different from Eqs.~(\ref{AFMEOM2}) derived on the basis of symmetry analysis in the absence of spin-orbit interaction. 

The very pronounced, highly anisotropic Gilbert damping terms in the Landau-Lifshitz-Gilbert equations of Eqs.~(\ref{AFMEOM3}) represent the main result of our paper. The phenomenon of the giant Gilbert damping anisotropy in the 2D AFM clearly calls for a qualitative understanding that we provide in Sec.~\ref{sec:qual}.

\section{Qualitative consideration}\label{sec:qual}

The results of Eqs.~(\ref{gilbert}), (\ref{coeffGD}) suggest that the anisotropy of Gilbert damping is most pronounced in the metal limit, $\ep_F\gg \Delta+\lambda$ as far as $\lambda \tau/\hbar \gg 1$. In particular, certain spin density responses are vanishing in this limit. One of them is the response of the average spin density $\delta s_z^+$ to $\dot{m}_z$ that is defined by the tensor component $\alpha^{zz}_m$ in Eq.~(\ref{response}). The other four vanishing tensor components $\alpha^{xx}_n$, $\alpha^{xy}_n$, $\alpha^{yx}_n$ and $\alpha^{yy}_n$ correspond to the responses of the in-plane staggered spin densities $\delta s_{x}^-$ and $\delta s_{y}^-$ to $\dot{n}_x$ and $\dot{n}_y$.  

It is important to stress that the component $\alpha^{zz}_m$ is not only finite but also quite large in the absence of spin-orbit interaction, i.\,e. for $\lambda=0$. It is, therefore, instructive to understand how the onset of spin-orbit interaction may cancel $\alpha^{zz}_m$ response and lead to the conservation of $z$ projection of magnetization vector.  

Such a qualitative understanding can be achieved by considering the Kubo-Greenwood formula for $\alpha^{zz}_m$ for the model 
of Eq.~(\ref{full}) in the limit $\Delta\to 0$ and $\tau\to \infty$, 
\be
\label{Kubo}
\alpha^{zz}_m \propto \s_{\bb{p}}\!\!\s_{s,s'=\pm} \!|\la \Psi_{\bb{p},s}| \sigma_z |\Psi_{\bb{p},s'}\ra |^2
\delta(\ep_\textrm{F}-\epsilon^{e}_{p,s})\delta(\ep_\textrm{F}-\epsilon^{e}_{p,s'}),
\e
where $\epsilon^e_{p,\pm}=\sqrt{v^2p^2+\lambda^2/4}\mp\lambda/2$ correspond to the two electronic branches of Eq.~(\ref{spectrume}) that are evidently valley degenerate in the limit $\Delta\to 0$. 

The states $\Psi_{\bb{p},s}$ are simply the eigenstates of the Hamiltonian $H_0 = v\bb{p}\cdot\bb{\Sigma}+(\lambda/2) \lt[\bb{\sigma}\times\bb{\Sigma}\rt]_z$,
\be
H_0 =
\bpm
0 & 0 & v(p_x \!-\!i p_y) & 0\\
0 & 0 & -i \lambda & v(p_x\!-\!i p_y) \\
v(p_x\!+\!i p_y) & i\lambda & 0 & 0\\
0 & v(p_x\!+\!i p_y) & 0 & 0
\epm,
\e
that can be explicitly written as
\be
\Psi_{\bb{p},\pm}=
\frac{1}{2\sqrt{v^2p^2\mp\lambda \epsilon^e_\pm/2}} \bpm vp\, e^{-i \phi}\\ \pm i \epsilon^e_\pm\\ \epsilon^e_\pm\\ \pm i v p\, e^{i \phi}\epm,
\e
where we have used $p_x=p\cos\phi$, $p_y=p\sin\phi$. 

One may notice that $\la \Psi_{\bb{p},s}| \sigma_z |\Psi_{\bb{p},s}\ra =0$ for any value of $\lambda$ suggesting that the response function $\alpha^{zz}_m$ in Eq.~(\ref{Kubo}) is vanishing. This is, however, not the case for $\lambda=0$. Indeed, in the absence of spin-orbit interaction the electron branches become degenerate $\epsilon^e_{p,\pm}=v p$ such that the in-plane spin-flip processes contribute to the Kubo formula,
\be
\lt.\la \Psi_{\bb{p},+}| \sigma_z |\Psi_{\bb{p},-}\ra\rt|_{\lambda=0}=\lt.\la \Psi_{\bb{p},-}| \sigma_z |\Psi_{\bb{p},+}\ra\rt|_{\lambda=0}=1.
\e
These processes are exactly the ones responsible for a finite Gilbert damping component $\alpha^{zz}_m$ in the absence of spin-orbit interaction. The spin-orbit induced splitting of the subbands  forbids these spin-flip processes as soon as $\lambda\tau/\hbar \gg 1$  and leads to a giant anisotropy of Gilbert damping in the metal limit. Indeed, the other elements of the Gilbert damping tensor $\alpha^{xx}_m$ and  $\alpha^{yy}_m$ remain finite irrespective of the subband splitting,
\be
\la \Psi_{\bb{p},\pm}| (\sigma_x+i\sigma_y) |\Psi_{\bb{p},\pm}\ra=\frac{\pm i v\,p\,e^{i\phi}}{\sqrt{v^2p^2+\lambda^2/4}}.
\e
One can further show that for $\lambda=0$ the entire Gilbert damping tensor $\hat{\alpha}_m$ becomes isotropic $\hat{\alpha}^{xx}_m=\hat{\alpha}^{yy}_m=\hat{\alpha}^{zz}_m$ as it have been expected on the basis of the symmetry analysis. 

Very similar physics is also responsible for the anisotropy of the tensor $\hat{\alpha}_n$. It is worth noting that the same type of anisotropy is known to take place in the limit of large spin-orbit interaction in 2D Rashba ferromagnets \cite{AdoSTTGD2019}. Spin-orbit induced anisotropy of Gilbert damping plays, however, a lesser role in 2D ferromagnets due to the much stricter constraint on the single magnetization vector.  A less restricted dynamics of $\bb{m}$ and $\bb{n}$ vectors make the Gilbert damping anisotropy play a bigger role in 2D AFMs. 

Indeed, it can be directly seen from Eqs.~(\ref{AFMEOM3}) that a nonequilibrium state with  $\bb{m}=m \hat{\bb{z}}$ and $\bb{n}=\bb{n}_\parallel$ becomes undamped in the absence of external field $\bb{H}=0$. Such a state corresponds to the undamped N\'eel vector precession around $\hat{\bb{z}}$ axis with a frequency given by $J_\textrm{ex} m$. The state clearly survives in the presence of easy plane magnetic anisotropy in the AFM. We believe that such a phenomenon remains to be rather generic for a variety of 2D or layered AFM systems with strong spin-orbit coupling of Rashba type. 

\section{Conclusions}

In this paper, we demonstrate that the presence of sufficiently strong spin-orbit coupling $\lambda\tau/\hbar \gg 1$ results in the ultimate anisotropy of the Gilbert damping tensor in the metal regime, $\ep_F\gg\Delta+\lambda$.  One can trace the phenomenon to the spin-orbit induced splitting of Fermi surfaces that forbids scattering processes that are responsible for the relaxation of certain magnetization and N\'eel vector components. 

We also demonstrate that a finite in-plane projection $\bb{n}_\parallel$ of the N\'eel vector is responsible for a weak anisotropy of conductivity and spin-orbit torques for Fermi energies approaching the band edge, $\ep_F \sim \Delta+\lambda$. This anisotropy is, however, absent in the metallic regime. 

Gilbert damping is, however, in the absence of spin-orbit interaction as it is required by symmetry considerations. Thus, we demonstrate that the onset of Rashba spin-orbit interaction in 2D or layered AFM systems leads to a giant anisotropy of Gilbert damping in the metallic regime. The physics of this phenomenon originates in spin-orbit induced splitting of the electron subbands that destroys a particular decay channel for magnetization and leads to undamped precession of the N\'eel vector. The phenomenon is based on the assumption that other Gilbert damping channels  (e.\,g. due to phonons) remain suppressed in the long magnon wavelength limit that we consider. The predicted giant Gilbert damping anisotropy  may have a profound impact on the N\'eel vector dynamics in a variety of 2D and layered AFM materials. 

\begin{acknowledgments}
We are thankful to I.\,Ado, H.\,Gomonay and J.\,Sinova for fruitful discussions. This research was supported by the JTC-FLAGERA Project GRANSPORT.  D.Y. and M.T. acknowledge the support from the Russian Science Foundation Project No. 17-12-01359. A.P. acknowledges support from the Russian Science Foundation Project 18-72-00058. The work of D.Y. was also supported by the Swedish Research Council (Vetenskapsr{\aa}det, 2018-04383). M.T. is especially thankful to the KITP visitor program ``Spintronics Meets Topology in Quantum Materials''. O.E. acknowledges support from the Swedish Research Council (Vetenskapsr{\aa}det) and the Knut and Alice Wallenberg foundation.
\end{acknowledgments}

\appendix

\section{Model system}\label{sec:appa}

In this Appendix, we shall briefly justify Eqs.~(\ref{eff}) and (\ref{full}) of the main text. We start from an $s$-$d$-like model for two-dimensional antiferromagnet on a honeycomb lattice \cite{Sumit2019}. The model includes a local exchange interaction between localized magnetic moments and conduction electron spins as given by Eq.~(\ref{ex}). Itinerant electrons in the model are, therefore, governed by the tight-binding Hamiltonian 
\begin{align}
H_0=\,&-t\sum\limits_{i}\s_{\sigma\sigma'}c_{i\sigma}^\dagger c_{i\sigma'}
-J \s_{i} \s_{\sigma\sigma'}\bb{S}_i\cdot \bb{\sigma}_{\sigma\sigma'}c^\dagger_{i\sigma}c\0_{i\sigma'}\n\\
&+\frac{i\lambda}{3a}\s_{\langle i,j\rangle}\sum\limits_{\sigma\sigma'}\hat{\bm{z}}\cdot(\bm{\sigma}\times\bm{d}_{ij})_{\sigma\sigma'}c_{i\sigma}^\dagger c_{j\sigma'},
\label{htb}
\end{align}
where we do ignore disorder for a moment. The model is characterized by the nearest neighbor hopping energy $t$ and the Rashba spin-orbit coupling energy $\lambda$, $z$-axis is aligned perpendicular to the two-dimensional plane, the in-plane vectors $\bm{d}_{ij}$ connect the neighboring sites $i$ and $j$ on a honeycomb lattice. For any site $i$ on the sublattice $A$ we choose
\be
\bb{d}_{1}= a \bpm 0 \\ 1 \epm, \quad \bb{d}_{2}= \frac{a}{2} \bpm \sqrt{3} \\ -1 \epm , \quad \bb{d}_{3}= -\frac{a}{2} \bpm \sqrt{3}  \\ 1 \epm,
\e
where $a$ is the length of the bond between $A$ and $B$.

By projecting the tight-binding model of Eq.~(\ref{htb}) on states in a vicinity of the valley wave-vectors,
\be
\bb{K}= \frac{4\pi}{3\sqrt{3} a}\bpm 1\\ 0\epm,\quad\mbox{and}\quad \bb{K}'= -\bb{K},
\e
we find, in the valley symmetric approximation, the effective Hamiltonian of Eq.~(\ref{eff}) with the assumption that $\bb{S}^A=-\bb{S}^B$, where $v = 3ta/2\hbar$. By relaxing the assumption we obtain the model of Eq.~(\ref{full}).

\section{Linear Response tensors}\label{sec:appb}
\label{app:vertexcorrections}
In order to keep technical expressions compact we let $\hbar=1$ and $\ep_F=\ep$ below. Our technical analysis is based on linear response of electron spin density to various perturbations at zero frequency ($dc$) limit. In particular, we consider three types of responses: the one with respect to electric current (via electric field and inverse conductivity tensor), the one with respect to the time derivative of the N\'eel vector 
and the other one with respect to the time derivative of magnetization vector. These responses are summed up as
\beml
\label{tensors}
\begin{align}
\delta\bb{s}^+=\hat{S}^\textrm{SOT}_+\bb{j} +\hat{S}^\textrm{GD}_{mn} \dot{\bb{n}}+\hat{S}^\textrm{GD}_m\dot{\bb{m}},\\
\delta\bb{s}^-=\hat{S}^\textrm{SOT}_-\bb{j}+\hat{S}^\textrm{GD}_{nm}\dot{\bb{m}}+\hat{S}^\textrm{GD}_n \dot{\bb{n}},
\end{align}
\eml
where we define the response tensors $\hat{S}^\textrm{SOT}_\pm$ that are describing spin-orbit torques (both field-like and anti-damping) and various $\hat{S}^\textrm{GD}$ tensors that are describing various contributions to Gilbert dampings (and to effective spin renormalizations) \cite{AdoSTTGD2019}.

In order to compute the linear response tensors in Eqs.~(\ref{tensors}) we apply the standard Kubo formula
\be
\label{eq:kubo}
\delta\bb{s}^\pm_\alpha = 
\frac{J^2Sv^2\mathcal{A}}{2 V}
\s_\beta \widehat{\tr}  \lt\la \hat{G}^\text{R} \hat{s}^\pm_\alpha \hat{G}^\text{A}  
\hat{F}_\beta \rt\ra \,\frac{\pa X_\beta}{\pa t},
\e
where $V$ is the system area, $\widehat{\tr}$ is an operator trace, $\hat{G}^\text{R(A)}=(\ep-H\pm i 0)$ are retarded (advanced) Green function operators, $\hat{s}_\alpha^+ = \sigma_\alpha$, $\hat{s}_\alpha^-=\Lambda_z\Sigma_z\sigma_\alpha$ are the operators corresponding to the average spin-polarization $\bb{s}^+$ and staggered spin-polarization $\bb{s}^-$, the product $\hat{\bb{F}}\cdot \bb{X}(t)$ represents the time-dependent perturbation in the Hamiltonian, while the angular brackets denote the disorder averaging that we consider in diffusive (ladder) approximation.

The linear-response formula Eq.~(\ref{eq:kubo}) assumes zero temperature and zero frequency limit that corresponds to taking both Green's functions at the same energy $\ep=\ep_F$. We also neglect the Fermi-sea contribution (also known as St\v{r}eda contribution) since such a contribution appears to be either zero or subleading in the metal parameter $\ep\tau\gg 1$ with respect to our results.  
 
Thus, in order to compute Gilbert dampings and spin-orbit torque tensors we consider linear response of $\delta\bb{s}^\pm$ to the three perturbations mentioned above. Each perturbation is parameterized by the term $\delta H=\hat{\bb{F}}\cdot \bb{X}(t)$ with
\beml
\label{eq:ops}
\begin{align}
&\dot{\bb{X}}=\dot{\bb{n}},\qquad \hat{\bb{F}}=-\Delta\,\Lambda_z\Sigma_z\bb{\sigma},\\ 
&\dot{\bb{X}}=\dot{\bb{m}},\qquad\qquad\hat{\bb{F}}=-\Delta\, \bb{\sigma},\\
&\dot{\bb{X}}=(\pi v/e)\hat{\sigma}^{-1}\bb{j},\quad \hat{\bb{F}} =\bb{\Sigma},
\end{align}
\eml
where $\hat{\sigma}$ is the conductivity tensor (this is computed from the standard Kubo formula which is analogous to the one in Eq.~(\ref{eq:kubo}) but for the response of current density to electric field). The disorder averaging amounts to replacing Green's functions in Eq.~(\ref{eq:kubo}) with the corresponding disorder-averaged Green's functions and to replacing one of the operators, $\hat{s}_\alpha$ or $\hat{F}$, with the corresponding vertex-corrected operator.

\begin{figure}
\centering
\includegraphics{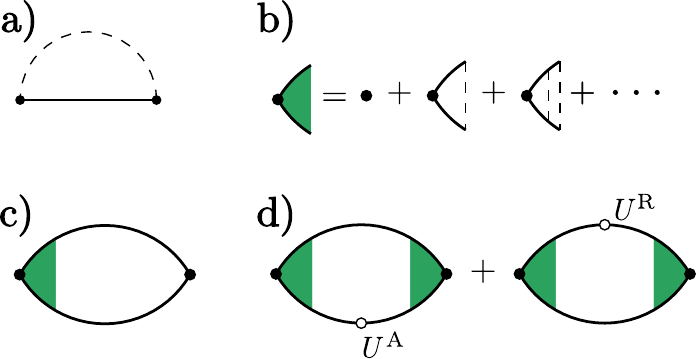}
\caption{Diagrammatic illustration. a) Born-approximation. b) Ladder-approximation. c) Disorder-averaged polarization bubble. d) Perturbative expansion of the disorder-averaged polarization bubble. }
\label{fig:diagrams}
\end{figure}

Disorder-averaged Green's functions become diagonal in the momentum space due to restored translational invariance and take the form 
$G^\text{R(A)}_{\bb{p}} = [\ep-H-\Sigma^\textrm{R(A)}]^{-1}$, where the Hamiltonian $H$ is defined in Eq.~(\ref{eff}) of the main text, while the self-energy $\Sigma^\text{R(A)}$ is evaluated in the Born-approximation depicted schematically in Fig.~\ref{fig:diagrams}a. 

We find that the real part of the self-energy does renormalize the Fermi energy $\ep$ and the $s$-$d$ exchange coupling strength $\Delta$, while the imaginary part reads
\be
\im \Sigma^\text{R(A)} = \mp\frac{\pi\alpha_d}{2}\,(\ep-\Delta\,\Lambda_z\Sigma_z\,\bb{n}\cdot\bb{\sigma}).
\e

In order to evaluate linear response tensors in the leading order with respect to the metal parameter $\ep\tau \gg 1$ one also needs to sum up the ladder diagrams as shown in Fig.~\ref{fig:diagrams}b-c.

To do that one defines the vertex corrected operator
\be
\label{eq:ladder}
\hat{F}^\text{vc}= \hat{F}+\hat{F}^{(1)} +\hat{F}^{(2)}+\hat{F}^{(3)}+\cdots,
\e
where we denote by $\hat{F}^{(i)}$ the operator $\hat{F}$ that is dressed by the number of $i$ disorder lines,
\be
\label{eq:onedisorderline}
\hat{F}^{(i)} = 2\pi\alpha_d\int\frac{\mathrm{d}^2\bb{p}}{(2\pi)^2} G_{\bb{p}}^\text{R}\hat{F}^{(i-1)}G_{\bb{p}}^\text{A}. 
\e
It appears that the summation in Eq.~(\ref{eq:ladder}) can be reduced to geometric series in a finite operator space. 
Indeed, let us define the operator space that is spanned by 16 operators in each of the valleys
\be
B_i=\frac{1}{2}\Lambda_\zeta\Sigma_\alpha \sigma_\beta,\quad i=\{\zeta,\alpha,\beta\},
\e
where $i$ is a cumulative index with $\zeta =0,z$ a valley parity index and $\alpha, \beta$ taking on the four values $\{0,x,y,z\}$ each. 

For $\bb{B}=(B_1, B_2,\dots, B_{32})$ we define the vertex corrected operator vector as 
\be
\label{sum}
\bb{B}^{\text{vc}}=\bb{B}+\mathcal{F}\bb{B}+\mathcal{F}^2\bb{B}+\mathcal{F}^3\bb{B}+\dots=
\frac{1}{1-\mathcal{F}}\bb{B},
\e
where $\mathcal{F}$ stands for a matrix of vertex corrections. Using the normalization condition $\tr B_i B_j =2\delta_{ij}$ 
we find
\be
\label{eq:matrixF}
\mathcal{F}_{ij} = \pi\alpha_d \int \frac{\mathrm{d}^2\bb{p}}{(2\pi)^2} 
\tr \lt[G^\text{A}_{\bb{p}} B_i G^\text{R}_{\bb{p}} B_j\rt],
\e
where $\tr$ stands for the usual matrix trace in the valley, spin and sublattice spaces.   

It easy to imagine that the matrix inversion in Eq.~(\ref{sum}) might be a daunting analytical task. We note, however, that the matrix $\mathcal{F}$ is evidently diagonal in the valley space, and it can also become block-diagonal in sublattice and spin spaces by choosing a more convenient basis. 

A particularly useful choice of basis corresponds to in-plane rotation of both spin and sublattice Pauli matrices to the frame associated with the in-plane projection $\bb{n}_\parallel$ of the N\'eel vector. For spin Pauli matrices this transformation is given by
\be
\label{trans}
\sigma_x \rightarrow \Lambda_z\frac{n_x \sigma_x + n_y \sigma_y }{\sqrt{n_x^2+n_y^2}}, \quad \sigma_y \rightarrow \Lambda_z\frac{n_y \sigma_x - n_x \sigma_y }{\sqrt{n_x^2+n_y^2}},
\e
where we took advantage of the fact that the direction of $\bb{n}$ is opposite in the two valleys. The same transformation (\ref{trans}) has to be applied to $\Sigma_{x}$ and $\Sigma_{y}$. 

The matrix $\mathcal{F}$ is instrumental for the analysis of all linear response tensors in Eq.~(\ref{tensors}). Indeed, using the definition of Eq.~(\ref{eq:matrixF}) in Eq.~(\ref{eq:kubo}) and summing up the diffusion ladders we find
\be
\label{final}
\delta s^\pm_\alpha = \frac{J^2S v^2\mathcal{A}}{2 \pi \alpha_d}
\s_\beta \s_{ij} \tr[\hat{s}^\pm_\alpha B_i] \mathcal{R}_{ij} \tr[\hat{F}_\beta B_j] \,\frac{\pa X_\beta}{\pa t},
\e
where $\mathcal{R}=\mathcal{F}(1-\mathcal{F})^{-1}$. Thus, the computation of all response tensors is reduced in the diffusive approximation to the computation of the vertex correction matrix $\mathcal{F}$ and subsequent matrix inversion.  

\section{Vertex correction}\label{sec:appc}

Still, finding an inverse matrix $(1-\mathcal{F})^{-1}$ is not that straightforward due to a pair of eigenvalues (one per valley) that equal exactly $1$. The presence of such eigenvalues roots in the particle conservation and is, therefore, not an artificial problem. The unit eigenvalues do evidently prevent the matrix inversion in Eq.~(\ref{sum}). Nevertheless, it can be shown that the corresponding eigenvectors do not enter the final equations of motion for localized spins. In the next section, we briefly illustrate how one can formally avoid the particle conservation divergence in the computation of vertex corrections.

Let us define by $\bb{a}_\zeta$ the eigenvectors of $\mathcal{F}$ that correspond to two unit eigenvalues, $\mathcal{F}\bb{a}_\zeta=\bb{a}_\zeta$, with $\zeta=0, z$. For the  normalized vector $\bb{a}_\zeta$ we define special operators
\be
\label{Bdiv}
\bar{B}_\zeta=\bb{a}_\zeta \cdot \bb{B} 
= \frac{\varepsilon-\Delta \Lambda_\zeta \Sigma_z\,\bm{n}\cdot\bm{\sigma}}{2\sqrt{\varepsilon^2+\Delta^2}},
\e
which are conserved with respect to impurity dressing $\bar{B}_\zeta=\bar{B}_\zeta^{(i)}$ for any order $i$. This means that the vertex corrected operator $\bar{B}_\zeta^\textrm{vc}$ is formally diverging in the $dc$ limit. In what follows, we formally write $\bar{B}_\zeta^{\text{vc}} = R_\infty \bar{B}_\zeta$, where the limit $R_\infty \to \infty$ is taken at the end of the calculation.

The response tensors defined by Eqs.~(\ref{response}) consist of different correlators of the operators $\Sigma_\alpha$, $s^+_\alpha=\sigma_a$, and $s^-_\alpha = \Lambda_z\Sigma_z\sigma_\alpha$. It is evident that most of these operators are already orthogonal to $\bar{B}_\zeta$,
\be
\tr \lt[\Sigma_\alpha \bar{B}_\zeta\rt] = \tr \lt[s^+_\alpha \bar{B}_\zeta\rt]= \tr \lt[s^-_\alpha \bar{B}_0\rt] =0,
\e
while the only dangerous sector is related to the projection
\be
\label{finite}
\tr \lt[s^-_\alpha \bar{B}_z\rt] = - \frac{4\Delta\, n_\alpha}{\sqrt{\varepsilon^2+\Delta^2}},
\e
which is evidently finite. The result of Eq.~(\ref{finite}) leads to formally diverging contribution $\delta \bb{s}^-_\textrm{div}$ that is generally present in all components of $\delta \bb{s}^-$, 
\be
\delta s^{-}_{\textrm{div},\alpha}\propto  R_\infty
\s_\beta \tr[\hat{\bb{s}}_\alpha^- \bar{B}_z] \tr[\hat{F}_\beta \bar{B}_z] \,\frac{\pa n_\beta}{\pa t}.
\e
One can immediately see, however, that such a diverging contribution corresponds to a particular vector form,
\be
\label{divergence}
\delta s^{-}_{\textrm{div},\alpha} \propto R_\infty n_\alpha\, \bb{n}\cdot\frac{\pa \bb{n}}{\pa t} =0,
\e
that manifestly vanishes due to the constraint $|\bb{n}|=1$ which is exact in the limit $\bb{m}=0$. Thus, the divergency in $B_\textrm{div}^\textrm{vc}$ (which originates in the diffusion pole of the density-density response) is, in fact, harmless for the response 
tensors we are discussing. 

It is interesting to note that the irrelevance of the divergency in $B_\textrm{div}^\textrm{vc}$ operator extends to higher orders in $\bb{m}$, even though it becomes much harder to see. We touch on this problem in Appendix~\ref{sec:appd}. 

\section{Finite magnetization}\label{sec:appd}
The deviation from a collinear antiferromagnetic order can be accounted by considering a finite net magnetization term in the Hamiltonain perturbatively,
\be
H =H^\textrm{eff}+ U, \qquad U = -\Delta\,\bb{m}\cdot\bb{\sigma}.
\e
In the paper, we build the first order perturbation theory with respect to $U$. 

First of all, it can be shown that the self-energy acquires the linear in $\bb{m}$ contribution as
\be\label{eq:SEcorr}
\im \Sigma^\text{R(A)} = \mp\frac{\pi\alpha_d}{2}\,(\ep-\Delta\,\Lambda_z\Sigma_z\,\bb{n}\cdot\bb{\sigma}+\Delta\,\bb{m}\cdot\bb{\sigma}).
\e
Second, the Dyson expansion of the disorder-averaged Green's functions $G^\text{R(A)}$ with respect to $\bb{m}$ reads
\be
G^\text{R(A)}\rightarrow G^\text{R(A)}+G^\text{R(A)}U^\text{R(A)}G^\text{R(A)},
\e
where $U^\text{R(A)}=U(1 \pm i \pi\alpha_d/2)$ and we disregarded terms starting from quadratic order in $\bb{m}$. Note, that 
we have kept the notations $G^\text{R(A)}$ for the disorder averaged Green's functions of the unperturbed system. 

The computation of linear response tensors amounts to considering an additional contribution to the response tensor represented by a complex class of diagrams depicted schematically in  Fig.~\ref{fig:diagrams}d. Before ladder summation is applied the diagrams of Fig.~\ref{fig:diagrams}d correspond to a contribution to the correlator of two operators $B_i$ and $B_j$ of the type 
\begin{align}
\mathcal{U}_{ij} =\,& 2\pi\alpha_d \int \frac{d^2\bb{p}}{(2\pi)^2} \tr\lt[G^\text{A} U^\text{A} G^\text{A} B_i G^R B_j\n\rt.\\
&\lt.+G^\text{A} B_i G^R U^R G^R B_j\rt],   
\end{align}
which has yet be dressed. The dressing amounts to replacing both $B_i$ and $B_j$ operators with the corresponding vertex corrected operators $B^\textrm{vc}_i$ and $B^\textrm{vc}_j$, respectively. 

The final result for the response of spin density is still given by Eq.~(\ref{final}), where the matrix $\mathcal{R}=\mathcal{F}(1-\mathcal{F})^{-1}$ is, however, replaced with 
\be
\label{eq:mresp}
\mathcal{R}=\frac{\mathcal{F}}{1-\mathcal{F}} + \frac{1}{1-\mathcal{F}}\mathcal{U}\frac{1}{1-\mathcal{F}},
\e
which corresponds to diagrams Fig.~\ref{fig:diagrams}c-d.
It is again convenient to consider a particular basis for the matrix $\mathcal{F}$ as defined in Eq.~(\ref{trans}) to simplify analytical computation.  

The problem of divergence in the operators $\bar{B}_\zeta$ does now become less trivial. Careful analysis shows that the linear terms in $\bb{m}$ included in Eq.~(\ref{eq:mresp}) lead to additional diverging contributions to $\delta\bb{s}^{-}$ of the form
\be
\label{extra}
\delta\bb{s}^{-, (1)}_{\text{div},\alpha} \propto  - R_\infty n_\alpha\, \bb{m}\cdot\frac{\pa \bb{m}}{\pa t},
\e
that is analogous to the one in Eq.~(\ref{divergence}) for a finite $m$. (We remind that the constraint $n^2+m^2=1$ provides a relation between these terms). The contribution in Eq.~(\ref{extra}) is, however, of too high order in $\bb{m}$ in Eq.~(\ref{ndot}) and cancels out completely in Eq.~(\ref{mdot}).

The terms linear in $\bb{m}$ are also responsible for diverging contributions in $\delta s_\alpha^{+}$ of the type 
\be
\delta s^{+}_{\text{div},\alpha} \propto R_\infty m_\alpha\, \bb{n}\cdot\frac{\pa \bb{n}}{\pa t}= - R_\infty m_\alpha\, \bb{m}\cdot\frac{\pa \bb{m}}{\pa t},
\e
that appear to be of higher than a linear order in $\bb{m}$, thus, exceeding our working precision.

Overall, one can show that the operators $\bar{B}_\zeta$ can be formally excluded by projecting the operator space of $B_i$ operators on the corresponding subspace. The latter is facilitated by the transformation $\mathcal{F}\to P\mathcal{F}P$, where 
\be
\label{PPP}
P= 1- \s_{\zeta=0,z}\bb{a}_\zeta \bb{a}\h_\zeta, 
\e
is the projection operator. Here, $\bb{a}$ stands for the column vector and $\bb{a}\h$ for the corresponding conjugated string vector. Eq.~(\ref{PPP}) facilitates the regularized computation of the vertex corrections and lead to the results presented in the paper.

\bibliographystyle{apsrev4-1}
\bibliography{Biblio}

\end{document}